\documentclass[]{tPHM2e}

\begin{document}
\doi{10.1080/14786435.20xx.xxxxxx}
\issn{1478-6443}
\issnp{1478-6435}
\jvol{00} \jnum{00} \jyear{2012} 

\markboth{S.L. Bud'ko and P.C. Canfield}{Philosophical Magazine}

\articletype{}

\title{Frequency dependence of the spin glass freezing temperatures in icosahedral R - Mg - Zn (R = rare earth) quasicrystals}

\author{Sergey L. Bud'ko and  Paul C. Canfield\\\vspace{6pt} {\em{Ames Laboratory U.S. DOE and Department of Physics and Astronomy, Iowa State University, Ames, Iowa 50011, USA}}\\\vspace{6pt}\received{ } }

\maketitle

\begin{abstract}

We present ac susceptibility measurements with the frequency spanning three orders of magnitude on single grain, icosahedral R - Mg - Zn (R = rare earth) quasicrystals. The freezing temperature in Gd-based, Heisenberg spin glasses in this family increases by $\sim 2\%$ with a frequency increase from 10 Hz to 10 kHz, whereas the freezing temperature in the non-Heisenberg members of the family is significantly more responsive to the frequency change (by 16 - 22 \%), suggesting that an additional magnetic anisotropy distribution in the non-Heisenberg spin glasses causes changes in the low frequency magnetic dynamics.

\bigskip

\bigskip

\end{abstract}

\section{Introduction}
Magnetic moment bearing, rare earth containing, quasicrystals, being an example of well ordered solids with sharp diffraction peaks but with conventional requirement of translational symmetry lifted, present a rare example of an "ideal" spin glass, in which the spin glass state probably arises from the multiplicity of the R - R distances in the quasicrystalline lattice, \cite{hat95a,hat95b,cha97a,fis99a,fis00a,can01a,seb04a} as opposed to a substitutional disorder in crystalline metallic spin glasses. \cite{wie00a} Successful growth of large, single grain, R$_9$Mg$_{34}$Zn$_{57}$ (R = rare earth) icosahedral quasicrystals \cite{fis98a,can01b} allowed for detailed studies of the physical properties of the spin glass state in these materials, leading, in particular, to a clear delineation of the experimental differences between Heisenberg and non-Heisenberg spin glasses. \cite{fis99a,fis00a,can01a} It was shown that the freezing temperature, $T_f$, is lower for the Gd-based, Heisenberg spin glasses, than for R = Tb - Er, non Heisenberg spin glasses for the samples with the same values of the de Gennes factor [$dG = (g-1)^2J(J+1)$] , or the Weiss temperature. As a consequence, for e.g. (Tb$_{1-x}$Gd$_x$)$_9$Mg$_{34}$Zn$_{57}$ and  (Dy$_{1-x}$Gd$_x$)$_9$Mg$_{34}$Zn$_{57}$ pseudo-ternary solid solution the maximum in $T_f$ was observed for $x \approx 0.7 - 0.8$  when a crossover from from Heisenberg to non-Heisenberg behavior occurs. \cite{can01a} 

Based on the large data sets in Refs. \cite{fis99a,fis00a,can01a} it was suggested that two factors give rise to the spin glass state in the  R$_9$Mg$_{34}$Zn$_{57}$ quasicrystals: distribution of R - R distances and distribution of easy axis (or easy plane) in the non-Heisenberg members of the family. Since (i) a Heisenberg quasicrystal Gd$_9$Mg$_{34}$Zn$_{57}$ has a spin glass low temperature state, \cite{can01a} and (ii)  an attempt to design spin glass just by mixing rare earths with different anisotropies in a crystalline structure has failed so far, \cite{law07a} the distribution of the R - R distances is apparently a {\it necessary} and {\it sufficient} condition for the formation of a spin glass, and distribution of magnetic anisotropies is neither. 

The frequency dependence of the freezing temperature (the position of the peak in ac susceptibility)  has been observed and discussed in a number of spin glass systems. \cite{dah79a,hol79a,tho80a,kau83a} The fractional relative change in the freezing temperature per decade of frequency, $\Delta T_f/[T_f \Delta (log_{10} f)]$ ($f$ is frequency) was noticed to vary by more than an order of magnitude for different spin glasses. \cite{myd93a} It was suggested \cite{tho80a} that the experimental data could be well described by the empirical Vogel-Fulcher law,
\[
f = f_0~exp \left [ \frac{-E_a}{k_B(T_f - T_0)} \right ]
\]
where $k_B$  is the Boltzmann constant, and $F_0$, $E_a$, and $T_0$ are the fitting parameters. From the very beginning, though, it was understood that the usual set of $T_f(f)$ experimental data would not be suitable for obtaining all three parameters, and the value of $f_0$ was either obtained from other measurements or estimated by some independent procedure (e.g. remnant magnetization measurements \cite{pre78a} were used to determine $f_0$ in \cite{tho80a}).

R$_9$Mg$_{34}$Zn$_{57}$ quasicrystals, with magnetic moment bearing rare earths give us an opportunity to address trends in the frequency dependence of the freezing temperature in the family of spin glasses with Heisenberg and non-Heisenberg members, a task that is rather difficult to undertake in dilute, substitutional, crystalline spin glasses. To the best of out knowledge frequency dependent measurements in this family were performed and analyzed so far only for Tb$_9$Mg$_{34}$Zn$_{57}$. \cite{fis99a}

\section{Experimental}
Large single grain R$_9$Mg$_{34}$Zn$_{57}$ quasicrystals were grown from ternary or pseudo-ternary melt as described in detail in Refs. \cite{fis98a,can01b}. The actual samples used in this work were taken from the batches extensively studied in the past.  \cite{fis99a,fis00a,can01a} The samples with R = Tb, Dy, Gd, Gd$_{0.8}$Y$_{0.2}$, Gd$_{0.6}$Y$_{0.4}$, and Gd$_{0.5}$Tb$_{0.5}$, were chosen for this work thus covering Heisenberg, diluted (disordered) Heisenberg, and non-Heisenberg spin glasses. Low temperature, low field (25 Oe) zero field cooled - warming and field cooled - warming dc susceptibility measurements were done in a Quantum Design MPMS-7 SQUID magnetometer so as to ensure that sample chosen indeed have a  low temperature spin glass state and are not rhombohedral approximants with a long range magnetic order. \cite{fis98a}. The low temperature ac susceptibility was measured in 3 - 5 Oe ac field at frequencies in the range of 10 Hz to 10 kHz,  in zero dc field using the ACMS option of Quantum Design PPMS-14 instrument. A criterion $d \chi'_{ac}/dT = 0$ ($\chi'_{ac}$ is the real part of the ac susceptibility) was used to infer a value for $T_f$.

\section{Results and Discussion}
\begin{figure}
\begin{center}
\resizebox*{10cm}{!}{\includegraphics{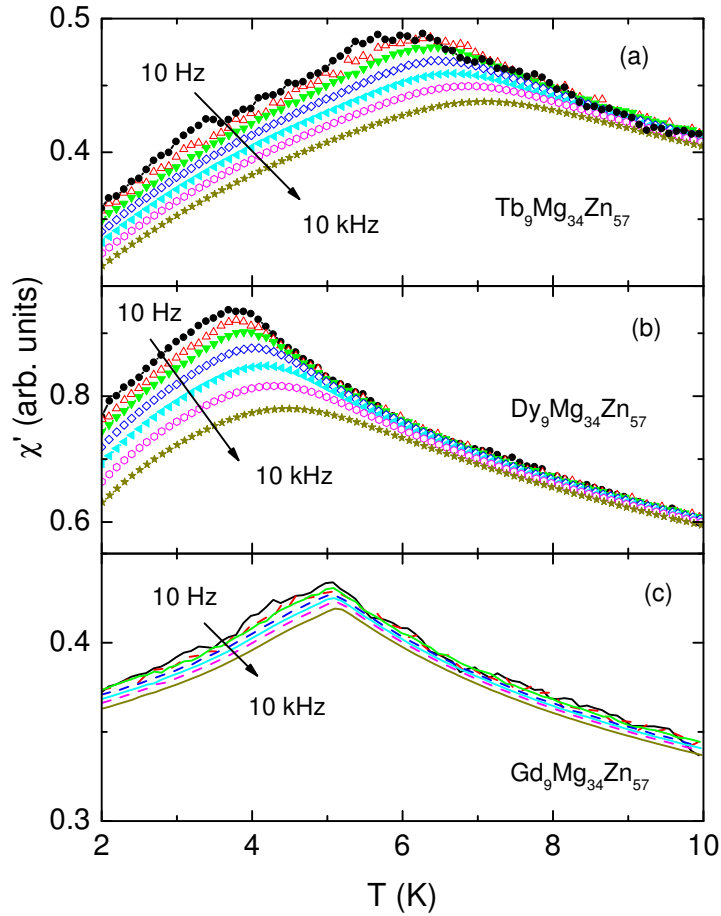}}%
\caption{Representative data for  R$_9$Mg$_{34}$Zn$_{57}$ (R = Tb, Dy, Gd) quasicrystals of the real part  of ac susceptibility measured using 10 Hz, 31.6 Hz, 100 Hz, 316 Hz, 1 kHz, 3.16 kHz and 10 kHz frequencies. The data for Gd$_9$Mg$_{34}$Zn$_{57}$ have the same data points density as those for Tb$_9$Mg$_{34}$Zn$_{57}$  and Dy$_9$Mg$_{34}$Zn$_{57}$. The lines were used in panel (c) for clarity.}%
\label{F1}
\end{center}
\end{figure}

Examples of low temperature ac susceptibility measurements are shown in Fig. \ref{F1}. The maximum in $\chi'_{ac}$ clearly shifts to higher temperatures at higher frequencies for the non-Heisenberg spin glasses Tb$_9$Mg$_{34}$Zn$_{57}$ and Dy$_9$Mg$_{34}$Zn$_{57}$ and is almost unchanged for the Heisenberg spin glass Gd$_9$Mg$_{34}$Zn$_{57}$.

\begin{figure}
\begin{center}
\resizebox*{10cm}{!}{\includegraphics{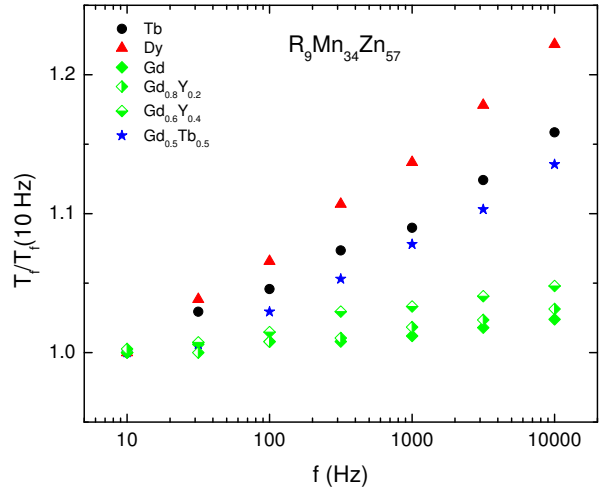}}%
\caption{Normalized spin glass freezing temperature, $T_f/T_f(10$ Hz$)$, as a function of frequency for ternary and preudo-ternary quasicrystals.}%
\label{F2}
\end{center}
\end{figure}

Fig. \ref{F2} summarizes all of such measurements performed in this work: whereas with three orders of magnitude frequency change, the $T_f$ of non-Heisenberg spin glasses (R = Tb, Dy) increases by about 20\%, the $T_f$ of the Heisenberg spin glass (R = Gd)  increases by mere $\sim2$\%. If substitutional disorder is added to the Heisenberg spin glass (R = Gd$_{0.8}$Y$_{0.2}$, Gd$_{0.6}$Y$_{0.4}$) the change is slightly larger, but still very much below those for the non-Heisenberg systems. For a mixture of Heisenberg and non-Heisenberg magnetic moments, R = Gd$_{0.5}$Tb$_{0.5}$, the relative change in the $T_f$ is in between the extreme cases but actually quite close to the R = Tb data. The fractional relative change in the freezing temperature per decade of frequency, $\Delta T_f/[T_f \Delta (log_{10} f)]$ results are summarized in the Table \ref{T1}. The values are within the broad range reported for different spin glasses. \cite{myd93a}

\begin{table}
  \tbl{Fractional relative change in $T_f$ per decade of frequency and Vogel - Fulcher fit parameters (for fixed $f_0 = 4 \times 10^7$ Hz \cite{fis99a}) in R$_9$Mg$_{34}$Zn$_{57}$ quasicrystals. }
{\begin{tabular}{@{}lccll}\toprule
   R  & $T_f(10$ Hz$)$ & $\Delta T_f/[T_f \Delta (log_{10} f)]$
         & $E_a/k_B$ [K]
         & $T_0$ [K]\\
\colrule
   Tb &6.12 & 0.051(2) & 14.8(4) & 5.31(5)  \\ 
   Dy & 3.65 & 0.073(2) & 11.2(3) & 3.11(3) \\
   Gd &5.01 & 0.007(1) & 2.1(3) & 4.88(5)  \\
   Gd$_{0.8}$Y$_{0.2}$ & 3.83 & 0.010(1) & 2.0(1) & 3.70(2) \\
   Gd$_{0.6}$Y$_{0.4}$ & 2.71 & 0.016(1) & 1.4(1) & 2.67(2) \\
   Gd$_{0.5}$Tb$_{0.5}$ & 6.79 & 0.046(3) & 15.0(2) & 5.90(2) \\
   \botrule
  \end{tabular}}
      \label{T1}
\end{table}

\begin{figure}
\begin{center}
\resizebox*{10cm}{!}{\includegraphics{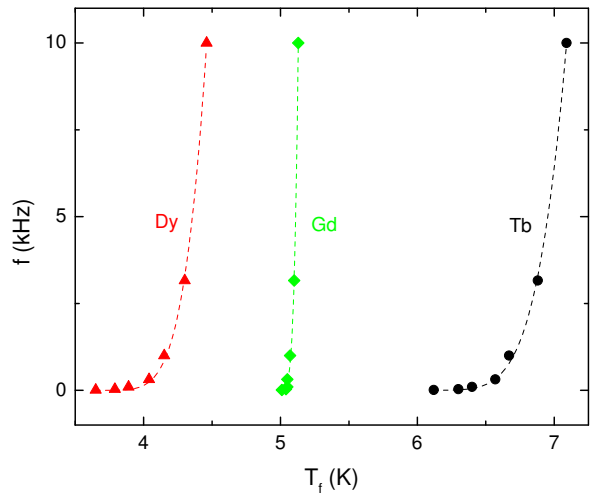}}%
\caption{$T_f$ - frequency data for R$_9$Mg$_{34}$Zn$_{57}$ quasicrystals with  R = Tb, Dy, and Gd. Dashed lines - the Vogel - Fulcher fits with $f_0 = 4 \times 10^7$ Hz.}%
\label{F3}
\end{center}
\end{figure}

The Volgel - Fulcher law can be used to fit the $T_f$ - frequency data. The fits are rather insensitive to the value of $f_0$. Here, for all samples, we have used $f_0 = 4 \times 10^7$ Hz obtained for Tb$_9$Mg$_{34}$Zn$_{57}$ \cite{fis99a}, although the values of $f_0$ as high as $10^{12}$ Hz give a comparable quality fits. Fits for R = Tb, Dy, and Gd are shown in Fig. \ref{F3} as an example.

The parameters of the fits for the ternary and pseudo-ternary quasicrystals in this work are listed in the Table \ref{T1}. For the non-Heisenberg spin glasses, $E_a/k_B > T_0$, as it was reported for a number of spin glasses. \cite{tho80a} For the Heisenberg and disordered Heisenberg spin glasses,  $E_a/k_B < T_0$, and the values are closer to each other. The significance of this difference is not clear at this point. It might just point out to some limitations of the Vogel - Fulcher fits for a small span of the freezing temperatures and a fixed value of $f_0$ throughout the family.

\section{Summary}

The systematic study of the frequency dependence of the spin glass freezing temperatures in the ternary and pseudo-ternary  R$_9$Mg$_{34}$Zn$_{57}$  quasicrystals revealed a distinct difference between non-Heisenberg and Heisenberg  members of the family, the latter showing significantly weaker response to the measurement frequency change (at least in the studied 10 Hz to 10 kHz range). It appears that the distribution of magnetic anisotropies (easy axis or plane) in the non-Heisenberg members changes the (low frequency) magnetic dynamics of these spin glasses. Similar trend was suggested (based on a very limited experimental data set) in crystalline RB$_{66}$ spin glasses. \cite{kim12a} It would be of interest to see if such distinct behavior is observed in other families of magnetic rare earth containing spin glasses, and if so, what theoretical models can be used or developed to account for such behavior.

\section*{Acknowledgments}

Help of I.R. Fisher, A.F. Panchula and K.O. Cheon in samples synthesis and decadal room temperature annealing sudies is greatly appreciated. Work at the Ames Laboratory was supported by the Department of Energy, Basic Energy Sciences, Division of Materials Sciences and Engineering under Contract No. DE-AC02-07CH11358. S.L.B. acknowledges partial support from the State of Iowa through Iowa State University.

\end{document}